\documentclass{article}
\usepackage[dvips]{graphicx}
\usepackage{epsfig}
\usepackage{multicol}

\author{Cecilia B. M. H. Chirenti}

\title{A numerical study on the dimension of an extremely
inhomogeneous matter distribution}

\date{\small{Instituto de F\'isica, Universidade de S\~ao Paulo\\
C.P.66.318, CEP 05315-970, S\~ao Paulo, Brazil}}

\begin{document}

\maketitle

\begin{abstract}
We have developed an algorithm that numericaly computes
the dimension of an extremely inhomogeneous matter distribution, given
by a discrete hierarchical metric. With our results it is possible to
analise how the dimension of the matter density tends to $d = 3\,$, as
we consider larger samples.
\end{abstract}

In a previous work \cite{Abdalla1}, we have presented a study on the
hierarchical metric  
\begin{equation}
dS^{2} = dt^{2} + g_{11}dx^{2} + g_{22}dy^{2} + g_{33}dz^{2} 
\ ,
\label{metric}
\end{equation}
where the metric is defined on all integers, depending on their
decomposition in terms of powers of 2 as  
\begin{eqnarray}
g_{11}(x) = a(t)^{2k}\,,\qquad &\textrm{with }&x=2^{k+1}n+2^{k}-1\nonumber\\
g_{22}(y) = a(t)^{2\ell}\,,\qquad &\textrm{with }&y=2^{\ell+1}n+2^{\ell}-1\\
g_{33}(z) = a(t)^{2m}\,,\qquad &\textrm{with }&z=2^{m+1}n+2^{m}-1 \,. \nonumber
\label{metric2}
\end{eqnarray}

We obtained the following expression for the matter density,
\begin{eqnarray}
T_{00} \equiv \rho = \frac{1}{8\pi G}
\frac{\dot{a}^{2}}{a^{2}}\left(k\ell+\ell m+mk\right)\nonumber\\ 
\equiv \rho_{0}(t)\left(k\ell+\ell m+mk\right)\,,
\label{tensor}
\end{eqnarray}
by means of the Einstein equations. Computing the Christoffel symbols
and subsequently the curvature tensor for this metric requires some
care, since we are not dealing with derivatives of functions, but
differences of functions defined on a discrete space.

We speculated whether such a matter distribution
could be described by a fractal. As it
turned out, from our preliminary analisis (see \cite{Abdalla1}), the
dimension of the matter density tends slowly to $d = 3$.

\begin{figure}[!hbp] 
\begin{center}
\epsfig{file=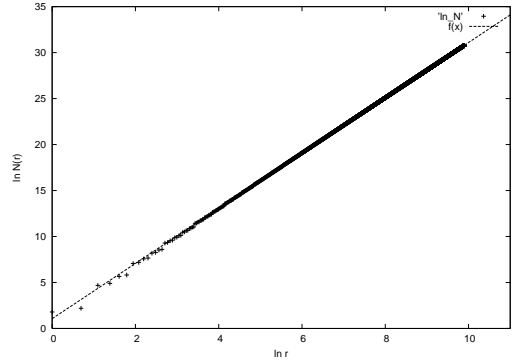,width=0.4\textwidth,angle=270,clip=}
\end{center}
\caption{\footnotesize{In this graphic we have $\ln N(r) \times \ln r$. The equation
of the straight line $f(x) = ax+b$ has \mbox{$a=3.00374 \pm 0.000025\,$},
 \mbox{$b=1.06288 \pm 0.000225\,$}.} }
\label{graph1}
\end{figure}

Considering the relation
\begin{eqnarray} 
\lim_{r \rightarrow \infty}
\frac{N(r)}{r^{d}} = K\,, \qquad
\textrm{with }N(r) = \sum_{0<x,y,z<r}\frac{\rho}{\rho_{0}}\,,
\label{lim}
\end{eqnarray}
where $K$ is some constant, and the constant $d$ is the fractal
dimension of the matter density \cite{Mandelbrot}, it is easy to see
that it implies
\begin{equation} 
\ln{N(r)} = \ln{K} + d\ln{r}\,,
\label{line}
\end{equation}
for large $r$. We have numerically computed $N(r)$ and the plot
$\ln{N(r)} \times \ln{r}$, in figure (\ref{graph1}), showing that equation
(\ref{line}), although valid at very large $r$), is a good
approximation for the behavior of the data.

\begin{figure}[!hbp]
\begin{center}
\epsfig{file=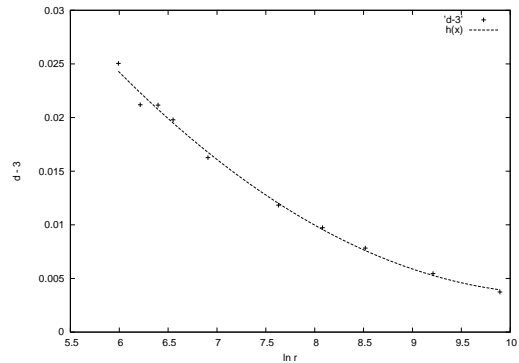,width=0.4\textwidth,angle=270,clip=}
\end{center}
\caption{\footnotesize{This graphic shows $(d-3)\times \ln r$, obtained for $r$ values
up to 20.000. The curve is a parabola fitted to the points by
Gnuplot, with equation \mbox{$h(x) = ax^{2}+bx+c\,$}:
$a = 0.00102 \pm 0.00014\,$,
\mbox{$b = -0.02134 \pm 0.00220\,$},
$c = 0.11576 \pm 0.008441\,$.}}
\label{graph2}
\end{figure}

Numerical fits for these points, taking increasingly larger samples,
have given results for $d$ that approach $d = 3$, as shown in
figure (\ref{graph2}).

As a conclusion, we can say, now, that the dimension of the matter
density that generates metric (\ref{metric}) tends to $d = 3$ as
shown in figure (\ref{graph2}).

\bigskip
{\bf Acknowledgements:}  This work has been supported by Funda\c c\~ao
de Amparo \`a Pesquisa do Estado de S\~ao Paulo {\bf (FAPESP)}, Brazil. 

\footnotesize{

}

\end{document}